\documentclass[aps,showpacs,twocolumn]{revtex4}
\usepackage{amssymb}
\usepackage{amsmath}
\usepackage{graphicx}
\usepackage{epsfig}

\begin{document}

\title{Fractional revivals of the quantum state in a tight-binding chain}
\author{Bing Chen$^{1}$, Z. Song$^{1,a}$ and C.P. Sun$^{1,2,a,b}$}
\affiliation{$^{1}$Department of Physics, Nankai University, Tianjin 300071, China}
\affiliation{$^{2}$ Institute of Theoretical Physics, Chinese Academy of Sciences,
Beijing, 100080, China}

\begin{abstract}
We investigate the time evolution of a Gaussian wave packet (GWP) in a
tight-binding chain with a uniform nearest neighbor (NN) hopping integral.
Analytical investigations and numerical simulations show that fractional
revivals of the quantum state occur in this system, i.e., at appropriate
times, a GWP can evolve into many copies of the initial state at different
positions. The application of this quantum phenomenon to quantum information
transfer in solid-state systems is discussed.
\end{abstract}

\maketitle

\section{Introduction}

Fractional revivals of a quantum state occur when the wave function evolves
in time to a state, at a specific instant between two full revivals, which
can be described by a superposition of states with equal amplitudes, each of
which has the same shape as that of the initial wave packet, yet with
different spatial distributions. The fractional revival is an interesting
phenomenon in quantum mechanics, which has no analog in classical physics.
It has been studied extensively, both theoretically \cite%
{Robinett,Averbukh,Aronstein,Agarwal,Vugalter} and experimentally \cite%
{Knospe,Vrakking,Lee,Spanner}. Most of the studies have been devoted to
continuous systems, such as the Coulomb potential and the infinite square
well. In this paper, we will focus on the phenomenon of fractional revival
in a discrete solid-state system. As an illustration, a simple tight-binding
chain with a uniform nearest neighbor (NN) hopping integral is investigated
analytically and numerically. Such a model is used to describe the Bloch
electronic system in condensed matter physics and now the qubit array
relevant to quantum information applications due to its equivalence with the
$XY$ spin chain. It is found that fractional revivals of a Gaussian wave
packet (GWP) occur in a discrete system. This opens up the possibility of
performing high-fidelity quantum information transfer (QIT) and creating
long-range entanglement by employing this novel feature of the time
evolution in this solid-state system.

Quantum information processing in solid-state systems has attracted
widespread attention because of the potential scalability of devices. Within
this context, quantum state transfer (QST) from one place to another in such
a system becomes a crucial issue and has been analyzed theoretically \cite%
{Bose1,Liying,Song,Alexandre,Karbach,Christandle1,Christandle2,Christandle3,Bose2,Bose3,Shitao1,Shitao2}%
. A great advantage of this approach is that no dynamical controls are
needed after one prepares the quantum state. In one of the pioneering works
\cite{Bose1}, Bose considered a regular one-dimensional spin chain with
Heisenberg interactions, which is able to transfer a quantum state over a
reasonable distance with the aid of a distillation process. Since then, a
number of interesting proposals have been made for quantum communication
through spin systems to improve the fidelity of the QST. One of them is to
choose the proper modulation of the coupling strengths as suggested in \cite%
{Christandle1,Christandle2,Christandle3}. In such a system, although an
arbitrary local quantum state will spread as the time evolves, after a
period of time the dispersed amplitudes will \textquotedblleft
refocus\textquotedblright\ at the receiving location of the chain. So
perfect state transfer can be realized. Another approach makes use of gapped
systems. The advantage of these schemes is that the intermediate spins are
only virtually excited. In this case, the two separated qubits are coupled
and realize the entanglement of two points \cite{Liying}. This ensures that
the transfer of the single-qubit state is achieved with a very high fidelity.

It should be pointed that QST and QIT are two different concepts. QST means
a local quantum state changes its location. It usually corresponds to the
translation or reflection of the initial wave function. On the other hand,
typical quantum information means the way (or mode) of the superposition of
two orthogonal states. A complete QST can achieve perfect QIT. However, it
is sufficient but not necessary.

So far, almost all the proposed schemes for QIT are based on the fact that a
properly designed qubit array can provide a unitary evolution operator $U(t)$
which can accomplish the task of transferring the local state $\left\vert
\psi _{A}\right\rangle $ at position $A$ to the target state $\left\vert
\psi _{B}\right\rangle $ located at $B$ via the process of time evolution $%
\left\vert \psi _{B}\right\rangle =U(t)\left\vert \psi _{A}\right\rangle $.

\begin{figure}[tbp]
\includegraphics[ bb=68 326 527 523, width=7 cm, clip]{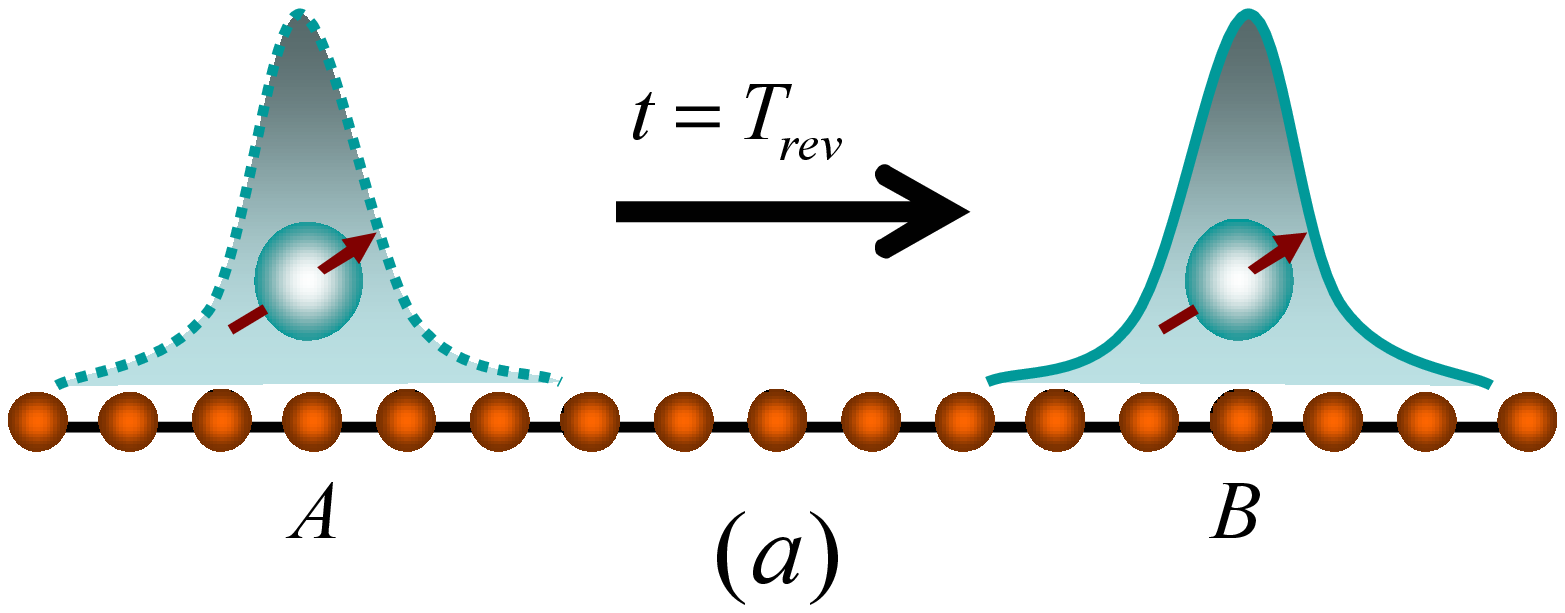} %
\includegraphics[ bb=39 178 505 587, width=7 cm, clip]{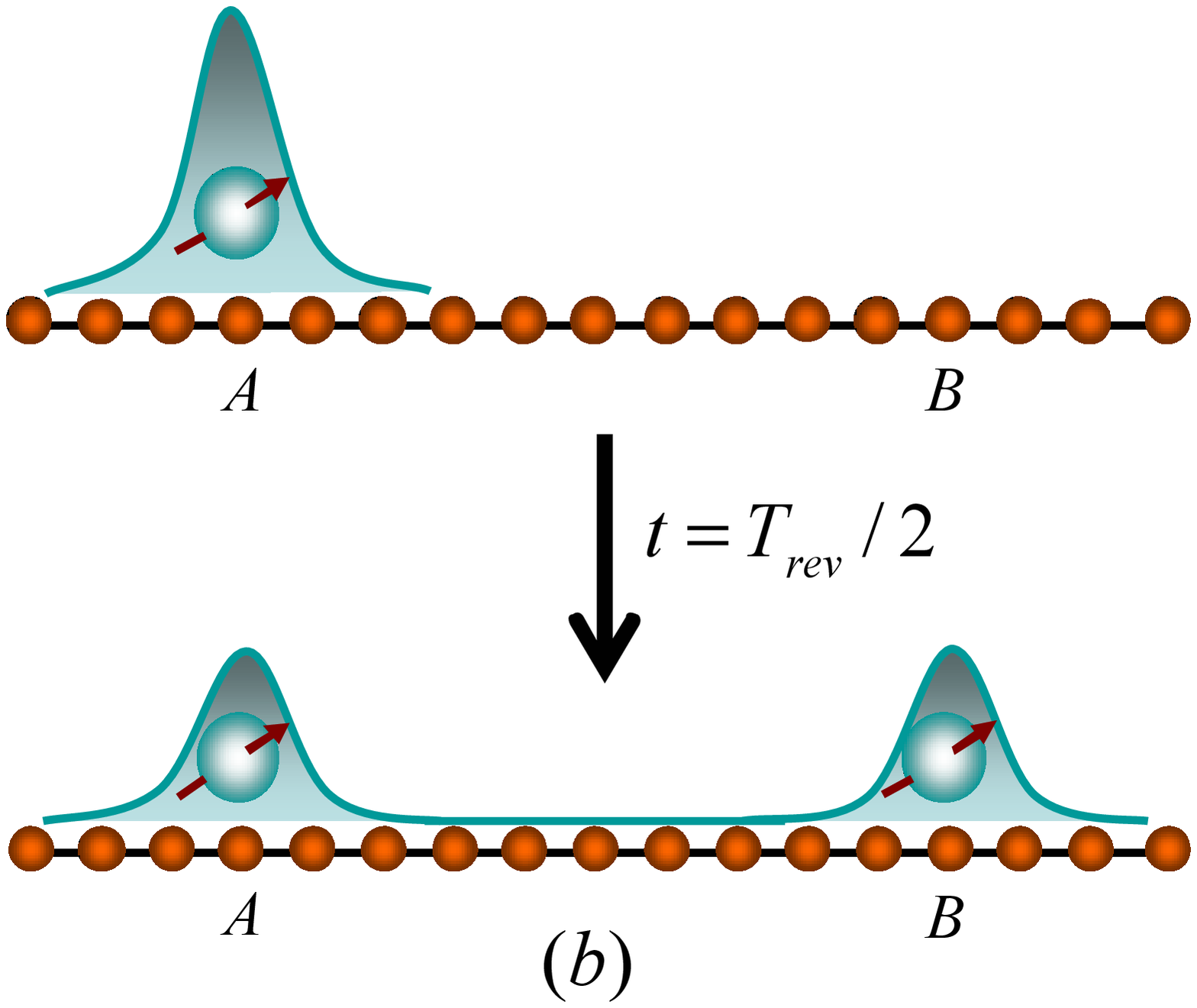}
\caption{(Color online) Schematic illustrations of quantum information
transfer implemented by the revivals of a polarized electronic wave packet:
(a) at instant $t=T_{rev}$, the initial wave packet at $A$ is in full
revival at $B$; (b) at instant $t=T_{rev}/2$, it is in half revival at $B$.}
\label{fig1}
\end{figure}

The phenomenon of quantum revival is an example of this process, which
transmits the \textit{complete} local state over the distance if $%
\left\langle \psi _{B}\right\vert U(t)\left\vert \psi _{A}\right\rangle =1$
(see Fig. \ref{fig1}a). In this case, we can say that perfect QST and QIT
are both achieved. However, theoretically, if the data bus guarantees the
\textit{partial} revival, i.e., $U(t)\left\vert \psi _{A}\right\rangle
=u_{B}\left\vert \psi _{B}\right\rangle +u_{C}\left\vert \psi
_{C}\right\rangle $ with $\left\langle \psi _{B}\right. \left\vert \psi
_{C}\right\rangle =0$ (see Fig. \ref{fig1}b), the quantum information
encoded in the target state $\left\vert \psi _{B}\right\rangle $\ can be
also extracted from the final state $U(t)\left\vert \psi _{A}\right\rangle $%
. In this case, the quantum state can not be transferred perfectly, but
perfect QIT is accomplished. We will show that such a scheme can be realized
based on the fractional revival of a quantum state.

In general, there are two ways to employ the tight-binding model as a data
bus for the QIT: (1) The qubit array is usually described by a Heisenberg
spin chain system \cite{Bose1}. Within the context of quantum state
transfer, only the dynamics of a single magnon are relevant. Thus in the
single-magnon invariant subspace, this model can be mapped into the single
spinless fermion tight-binding model. The quantum information is encoded in
the superposition of the single and zero particle states. (2) On the other
hand, the quantum information can also be encoded in the polarization of the
Bloch electron (Fig. \ref{fig1}). If the spin state of the Bloch electron is
a conserved quantity for the Hamiltonian of the medium, the spin state
cannot be influenced during the propagation, no matter how the spatial shape
of the wave function changes \cite{YS}. In this case, the locality of the
final state is crucial. We will discuss the two schemes in detail in section
IV based on the formalism of the fractional revival in the tight-binding
model.

%%%%%%%%%%%%%%%%%%%%%%%%%%%%%%%%%%%%%%%%%%%%%%%%%%%%%%%%%%%%%%%%%
%%%%%%%%%%%%%%%%%%%%%%%%%%%%%%%%%%%%%%%%%%%%%%%%%%%%%%%%%%%%%%%%%

This paper is organized as follows. In Sec. II the model is presented. In
Sec. III we formulate the general theory for the fractional revivals of a
GWP in the tight-binding model. In Sec. IV we show numerical results that
substantiate the analytical results. Sec. V is devoted to the application of
the theory for QST and QIT. Sec. VI is the summary and discussion.

\section{Model Hamiltonian}

For completeness, we consider a non-interacting spin-$1/2$ fermion system on
a one-dimensional lattice, which is a simple tight-binding $N$-site chain
with uniform NN hopping integral $-J$. The Hamiltonian can be written as
\begin{equation}
H_{e}=-J\sum_{j=1;\sigma =\uparrow ,\downarrow }^{N-1}\left( c_{j,\sigma
}^{\dagger }c_{j+1,\sigma }+h.c.\right) ,  \label{H}
\end{equation}%
where $c_{j,\sigma }^{\dagger }$ denotes the fermion creation operator at $j$%
-th site with spin $\sigma =\uparrow ,\downarrow $. Because this Hamiltonian
does not contain a spin-spin interaction, the polarization of the spin of an
electron is not changed as time evolves. Then the problem about the transfer
of the spin state can be reduced to the issue of charge transfer. In this
sense, we can concentrate on the spinless fermion model

\begin{equation}
H=-J\sum_{j=1}^{N-1}\left( a_{j}^{\dagger }a_{j+1}+h.c.\right) ,  \label{H1}
\end{equation}%
where $a_{j}^{\dagger }$ denotes the spinless fermion creation operator at $%
j $-th site, and the open boundary condition is applied. In section V, the
spin degree of freedom will be reconsidered for the discussion of QIT.
Introducing the Fourier-transformed fermion operator
\begin{equation}
\widetilde{a}_{k}^{\dagger }=\sqrt{\frac{2}{N+1}}\sum_{j=1}^{N}\sin
(kj)a_{j}^{\dagger },  \label{Fourior}
\end{equation}%
where $k=n\pi /\left( N+1\right) $, $n=1$, $2$, $...$, $N$, the Hamiltonian (%
\ref{H1}) can be diagonalized as
\begin{eqnarray}
H &=&\sum_{k}\epsilon _{k}\widetilde{a}_{k}^{\dagger }\widetilde{a}_{k}, \\
\epsilon _{k} &=&-2J\cos k,  \notag
\end{eqnarray}%
with the single-particle eigenstate
\begin{equation}
\left\vert \widetilde{k}\right\rangle =\sqrt{2/(N+1)}\sum_{j=1}^{N}\sin
(kj)\left\vert j\right\rangle ,  \label{standing}
\end{equation}%
where $\left\vert \widetilde{k}\right\rangle =\widetilde{a}_{k}^{\dagger
}\left\vert 0\right\rangle $, $\left\vert j\right\rangle =a_{j}^{\dagger
}\left\vert 0\right\rangle $.

In the large-$N$ limit, a discrete coordinate system approaches a continuous
one. Correspondingly, state (\ref{standing}) is a standing wave which is the
eigen wave function of the infinite square well. What is more, in lower
energy region $k\sim 0$,\ the spectrum is $\epsilon _{k}\approx $\ $%
-2J(1-k^{2})$\ $\sim 2Jk^{2}$, which is very close to that of the infinite
square well. This indicates that for low energies, the physics of the
tight-binding chain is approximately the same as that of the infinite square
well. Accordingly, for an initial state which can be expanded by the lower
eigenstates, its time evolution should be similar to that of the infinite
square well approximately, which is the typical paradigm to illustrate
fractional revivals \cite{Aronstein}. It implies that, for a wave packet
with low energy, the well-defined fractional revival formalism should be
valid in such a discrete coordinate system.

\section{Fractional revivals of GWP}

In this section, we investigate the time evolution of a GWP in the
tight-binding system analytically. Although the formalism presented in this
paper has been well established for continuous systems, we derive it exactly
to show that this formalism can be extended to discrete systems for some
special states such as GWPs.

\subsection{General formalism}

We consider an initial state in the low-energy range, which can be expanded
by the lower eigenstates. A typical state meeting this condition is the
zero-momentum GWP, which can be expressed as
\begin{equation}
\left\vert \psi \left( N_{0}\right) \right\rangle =\frac{1}{\sqrt{\Omega _{1}%
}}\sum_{j=1}^{N}e^{-\alpha ^{2}\left( j-N_{0}\right) ^{2}/2}\left\vert
j\right\rangle ,  \label{GWP1}
\end{equation}%
where $\Omega _{1}=\sum_{j=1}^{N}e^{-\alpha ^{2}\left( j-N_{0}\right) ^{2}}$
is the normalization factor; $N_{0}$ denotes the center of the GWP and $%
1+\Delta /2<N_{0}<$ $N-\Delta /2$ ensures that the entirety\ of the GWP is
situated within the chain approximately. The factor $\alpha $\ determines
the half-width
\begin{equation}
\Delta =\frac{2\sqrt{\ln 2}}{\alpha }
\end{equation}%
of the GWP in real space and also the range of the spectrum related to its
eigenstate expansion.

Firstly, we study the revival of a GWP in the framework of the
spectrum-parity matching condition (SPMC) \cite{Shitao2, LY2}, which is the
basis for the investigation of the fractional revival. Actually, using the
Fourier transformation (\ref{Fourior}), the GWP (\ref{GWP1}) can be written
as
\begin{equation}
\left\vert \psi \left( N_{0}\right) \right\rangle =\frac{1}{\sqrt{\Omega _{2}%
}}\sum_{k}\sin (kN_{0})e^{-k^{2}/2\alpha ^{2}}\left\vert \widetilde{k}%
\right\rangle ,  \label{GWP2}
\end{equation}%
where $\Omega _{2}=\sum_{k}\sin ^{2}(kN_{0})e^{-k^{2}/\alpha ^{2}}$ is the
normalization factor. For a system which has mirror symmetry, we have$\
[P,H]=0$, where $P$ is the reflection operator defined as $P\left\vert
j\right\rangle =\left\vert N+1-j\right\rangle $. For the model Hamiltonian (%
\ref{H1}), which has the mirror symmetry, the eigenstate $\left\vert
\widetilde{k}\right\rangle $\ satisfies $P\left\vert \widetilde{k}%
\right\rangle =p_{k}\left\vert \widetilde{k}\right\rangle $ with $p_{k}=\pm
1 $. So the mirror counterpart of the GWP (\ref{GWP1}) has the form
\begin{eqnarray}
P\left\vert \psi \left( N_{0}\right) \right\rangle &=&\left\vert \psi \left(
N+1-N_{0}\right) \right\rangle \\
&=&\frac{1}{\sqrt{\Omega _{2}}}\sum_{n}\left( -1\right) ^{n+1}\sin
(kN_{0})e^{-k^{2}/2\alpha ^{2}}\left\vert \widetilde{k}\right\rangle ,
\notag
\end{eqnarray}%
with the relation $k=n\pi /\left( N+1\right) $. For the GWP (\ref{GWP1})
with a large enough $\alpha $, it can be expanded by the eigenstate $%
\left\vert \widetilde{k}\right\rangle $\ with the eigenvalue $\varepsilon
_{k}$ and parities $p_{k}$ in the following way:
\begin{equation}
\varepsilon _{k}=n^{2}\Delta E,\text{ }p_{k}=(-1)^{n^{2}},  \label{ep}
\end{equation}%
where $\Delta E=2J\pi ^{2}/\left( N+1\right) ^{2}$ is the greatest common
divisor of all the possible level differences. Obviously,\ the dispersion
relation and the corresponding parity (\ref{ep}) satisfy the SPMC, which
leads to the following conclusion: for an initial state $\left\vert \phi
\left( N_{0},0\right) \right\rangle =\left\vert \psi \left( N_{0}\right)
\right\rangle $, at the instant $t=T_{rev}=\pi /\Delta E$, it evolves into
\begin{equation}
\left\vert \phi \left( N_{0},\frac{\pi }{\Delta E}\right) \right\rangle
=P\left\vert \phi \left( N_{0},0\right) \right\rangle .
\end{equation}%
Here we define $T_{rev}$ as the revival time, at which the wave packet
revives as the mirror reflection of the initial one. This is the same as in
an infinite square well, and is usually called a full revival.

In the following, we will show that fractional revival of the GWP also
occurs in the tight-binding model. We start our investigation with the time
evolution of the GWP $\left\vert \psi \left( N_{0}\right) \right\rangle $ in
the system (\ref{H1}). At the time $t$, the initial state $\left\vert \phi
\left( N_{0},0\right) \right\rangle $ has evolved into
\begin{eqnarray}
\left\vert \phi \left( N_{0},t\right) \right\rangle &=&e^{-iHt}\left\vert
\phi \left( N_{0},0\right) \right\rangle \\
&\simeq &\frac{e^{i2Jt}}{\sqrt{\Omega _{2}}}\sum_{k}\sin
(kN_{0})e^{-k^{2}/2\alpha ^{2}-i2Jk^{2}t}\left\vert \widetilde{k}%
\right\rangle .  \notag
\end{eqnarray}%
Here, we take $H\simeq -2J\sum_{k}(1-k^{2})\widetilde{a}_{k}^{\dagger }%
\widetilde{a}_{k}$ as the small-$k$ approximation for GWP. Neglecting the
overall phase $e^{i2Jt}$, we have
\begin{equation}
\left\vert \phi \left( N_{0},t\right) \right\rangle \simeq \frac{1}{\sqrt{%
\Omega _{2}}}\sum_{k}\sin (kN_{0})e^{-k^{2}/2\alpha
^{2}-i2Jk^{2}t}\left\vert \widetilde{k}\right\rangle .
\end{equation}%
The shape of $\left\vert \phi \left( N_{0},t\right) \right\rangle $\ in real
space as the result of the interference of the standing waves $\left\vert
\widetilde{k}\right\rangle $ depends on the time $t$\ and $N_{0}$. In
general, the feature of revival is characterized by the autocorrelation
function $A(t)=\left\langle \phi \left( N_{0},0\right) \right\vert \left.
\phi \left( N_{0},t\right) \right\rangle $. In this paper, we choose another
quantity, the mirror fidelity
\begin{equation}
F(t)=\left\langle \phi \left( N_{p0},0\right) \right\vert \left. \phi \left(
N_{0},t\right) \right\rangle  \label{ft}
\end{equation}%
where $N_{p0}=N+1-N_{0}$ denotes the mirror position of $N_{0}$,\ since the
revival of the wave packet at different locations is desirable for the task
of quantum state transfer and the generation of entanglement. The fidelity
is a quantity to characterize the QST.

Now we focus on the special instants $\tau =pT_{rev}/q$,\ where $p$, $q$\
are two mutually prime integers. In the framework of quantum information, $%
F(\tau )=1$\ indicates perfect quantum state transmission. At the moment $%
\tau $, the fidelity of QST can be expressed explicitly%
\begin{equation}
\left\vert F(\tau )\right\vert =\left\vert \sum_{n}\left( -1\right)
^{n+1}\left\vert a_{n}\right\vert ^{2}e^{-ipn^{2}\pi /q}\right\vert ,
\end{equation}%
where $a_{n}=\sqrt{1/\Omega _{2}}\sin (kN_{0})e^{-k^{2}/2\alpha ^{2}}$ is
the expansion coefficient, which is the starting point of our discussion.
Obviously, when $p=q$, we have $\tau =T_{rev}$ and $F(\tau )=1$. This result
is in agreement with the prediction from the SPMC, and can be employed to
perform the perfect QIT. As discussed in the introduction, it may be
possible to accomplish QIT in the case that the quantum state is not
transferred completely.

In the following, we will demonstrate it in the framework of the
well-defined formalism of fractional revivals. It is easy to see that $\exp
(-ipn^{2}\pi /q)$ is a periodic function with period $l$, i.e.%
\begin{equation}
e^{-ip\left( n+l\right) ^{2}\pi /q}=e^{-ipn^{2}\pi /q},
\end{equation}%
where $l$ is determined by $q$
\begin{equation}
l=\left\{
\begin{array}{c}
2q\text{ (odd\ }q\text{)} \\
q\text{\ (even }q\text{)}%
\end{array}%
.\right.  \label{period}
\end{equation}%
Performing the Fourier transformation, we have%
\begin{equation}
e^{-ipn^{2}\pi /q}=\sum_{r=0}^{l-1}b_{r}e^{-i2n\pi r/l},
\end{equation}%
where%
\begin{equation}
b_{r}=\frac{1}{l}\sum_{n=0}^{l-1}e^{i\left( 2n\pi r/l-pn^{2}\pi /q\right) }.
\end{equation}%
A straightforward calculation shows that $b_{r}$ satisfies the relation%
\begin{equation}
b_{r}=b_{l-r},\text{ }(r=1,2,\ldots ,l/2-1).
\end{equation}%
Then at the instant $\tau $, the state $\left\vert \phi \left(
N_{0},0\right) \right\rangle $\ evolves into%
\begin{eqnarray}
\left\vert \phi \left( N_{0},\tau \right) \right\rangle &=&\frac{1}{\sqrt{%
\Omega _{2}}}\sum_{n}a_{n}\sum_{r=0}^{l-1}b_{r}e^{-i2n\pi r/l}\left\vert
k\right\rangle  \notag \\
&=&b_{0}\left\vert \psi \left( N_{0}\right) \right\rangle
-b_{l/2}P\left\vert \psi \left( N_{0}\right) \right\rangle  \label{sub} \\
&&+\sum_{r=1}^{l/2-1}b_{r}[\left\vert \psi \left( N_{r}^{+}\right)
\right\rangle +\left\vert \psi \left( N_{r}^{-}\right) \right\rangle ],
\notag
\end{eqnarray}%
where $N_{r}^{\pm }=N_{0}\pm 2\left( N+1\right) r/l$. It is clear that, at
the time $\tau =pT_{rev}/q$, the initial GWP at $N_{0}$\ has evolved into $l$
sub-GWPs. Each of them closely reproduces the shape of the initial one but
at the positions $N_{p0}$, $N_{r}^{\pm }$ ($r=0$, $1$, $2$, $...$, $l/2-1$)
and the probability of the sub-GWP at $N_{r}^{\pm }$ $(N_{p0})$ is $%
\left\vert b_{r}\right\vert ^{2}$ $(\left\vert b_{l/2}\right\vert ^{2})$.
Eq. (\ref{sub}) shows that the evolved state at time $\tau $\ is the
superposition of as many as $l$\ \textquotedblleft clones\textquotedblright\
of the initial wave packet with amplitude $b_{r}$\ $(r=0$, $1$, $\cdots $, $%
l/2-1)$\ and different spacial positions. The final state $\left\vert \phi
\left( N_{0},\tau \right) \right\rangle $ is formed in the following
process. First, split the original wave packet into $l$\ copies of the
initial wave packet, each with the probability $\left\vert b_{r}\right\vert
^{2}$. Then translate the copies to the positions $N_{s}^{\pm }=N_{0}\pm
2\left( N+1\right) s/l$, $(s=0,1,...,l/2)$. For the sub-GWPs with $%
N_{s}^{\pm }$\ beyond the chain, i.e., $N_{s}^{+}>N$ or $N_{s}^{-}\leq 0$,
we have $\left\vert \psi \left( N_{s}^{+}\right) \right\rangle =e^{i\pi
}P\left\vert \psi \left( N_{s}^{+}-N\right) \right\rangle $\ and $\left\vert
\psi \left( N_{s}^{-}\right) \right\rangle =e^{i\pi }\left\vert \psi \left(
\left\vert N_{s}^{-}\right\vert \right) \right\rangle $\ from (\ref{GWP2}).
This indicates that $\left\vert \psi \left( N_{s}^{\pm }\right)
\right\rangle $ will be reflected with $\pi $-shift when $N_{s}^{\pm }$\ is
beyond the chain, and there are always two sub-GWPs at the initial position $%
N_{0}$\ and its mirror place $N_{p0}$ respectively.

Defining the function
\begin{equation}
f(N_{A},N_{B})=\left\langle \phi \left( N_{A},0\right) \right\vert \left.
\phi \left( N_{B},0\right) \right\rangle ,
\end{equation}%
thus the corresponding fidelity is
\begin{eqnarray}
F(\tau ) &=&\left\langle \phi \left( N_{p0},0\right) \right\vert \left. \phi
\left( N_{0},\tau \right) \right\rangle \\
&=&-b_{l/2}+\sum_{r=1,\lambda =\pm }^{l/2-1}b_{r}f(N_{p0},N_{r}^{\lambda }).
\notag
\end{eqnarray}%
Here, term $\left\langle \phi \left( N_{0},0\right) \right\vert \left. \phi
\left( N_{p0},0\right) \right\rangle $ has been ignored since we are only
interested in the case with $\left\vert N_{p0}-N_{0}\right\vert \gg \Delta $%
. Furthermore, from the relation%
\begin{equation}
b_{r}=e^{-i2\pi \left( r/l+p/q\right) }b_{r^{\prime }}
\end{equation}%
where $r^{\prime }=r+2pl/q$, we have%
\begin{equation}
\left\vert b_{r}\right\vert ^{2}=\frac{1}{q},  \label{br}
\end{equation}%
i.e., the probability of each sub-packet is $1/q$. However, Eq. (\ref{sub})
shows that the final $l$ sub-GWPs may be not orthogonal due to the
reflection and then their superposition determines the shape of the final
state. On the other hand, the overlap of two neighbor GWPs should affect the
shape of the final state if the sub-GWPs are too \textbf{numerous}. We will
discuss this in Sec. V with the aid of numerical simulation. Actually, the
above conclusion is also valid for an arbitrary initial state which
satisfies the low-energy condition.

\subsection{Application of the formalism}

In the following, our aim is to apply the formalism developed above to
several special cases, which should be useful for the transmission of
quantum information. From the above analysis, we have $|F(\tau )|=1$\ at $%
\tau =T_{rev}$, which shows that the initial GWP can be revived completely
at this instant. However, according to the formalism, $T_{rev}$ is not the
shortest period to perform perfect state transfer. If a proper $N_{0}$ is
chosen, a shorter period can be obtained. Here we investigate the case with $%
N_{0}=N/3$ to illustrate this point. When we consider the case $p/q=1/3$,
the corresponding period of the Fourier transformation in Eq. (\ref{period})
is $l=6$. Then at time $\tau =T_{rev}/3$, we have
\begin{equation}
\left\vert \phi \left( \frac{N}{3},\frac{T_{rev}}{3}\right) \right\rangle
=e^{-i\pi /3}P\left\vert \psi \left( \frac{N}{3}\right) \right\rangle .
\label{N/3}
\end{equation}%
It shows that at $\tau =T_{rev}/3$ the initial GWP recurs totally at its
mirror \textbf{counterpart}.

This result can also be explained in the framework of the SPMC. Actually,
for the eigenstate expansion of the initial GWP $\left\vert \psi \left(
N/3\right) \right\rangle $, it is easy to find that the expansion
coefficients of the levels $n=3m$ ($m$ is the integer) vanish. Then the
greatest common divisor of the effective levels for the state $\left\vert
\psi \left( N/3\right) \right\rangle $ is $\Delta E_{eff}=$ $6J\pi
^{2}/\left( N+1\right) ^{2}$ $=3\Delta E$, and the corresponding recurrence
period $T_{rev}^{\prime }=T_{rev}/3$. A similar analysis can be applied to
the case of the initial state $\left\vert \psi \left( N/2\right)
\right\rangle $. In this case, the expansion coefficients of the levels $%
n=2m $ vanish, and the corresponding recurrence period $T_{rev}^{\prime
}=T_{rev}/8$.\

Except these special cases, the greatest common divisor for the effective
levels of all GWPs at other positions remains $\Delta E$. Nevertheless, if
the number of the levels, which determines the greatest common divisor to be
$\Delta E$\ is few, an \textquotedblleft effective\textquotedblright\
greatest common divisor should govern the recurrence time dominantly. For
example, if we take $N_{0}=N/m$, $\left( m>3\right) $, we also notice large
partial revivals at $\tau =T_{rev}/m$. For $N_{0}=N/4$, at $\tau =T_{rev}/4$
the wave packet evolved into
\begin{eqnarray}
&&\left\vert \phi \left( \frac{N}{4},\frac{T_{rev}}{4}\right) \right\rangle
\notag \\
&=&\left( b_{0}-b_{1}\right) \left\vert \psi \left( \frac{N}{4}\right)
\right\rangle +\left( b_{1}-b_{2}\right) P\left\vert \psi \left( \frac{N}{4}%
\right) \right\rangle ,
\end{eqnarray}%
which represents four cloned small GWPs at positions $N/4$ and $3N/4$. Note
that the superposition leads to $|b_{0}-b_{1}|^{2}$ $=\left( 2-\sqrt{2}%
\right) /4$ $\approx 0.146$\ and $\left\vert b_{1}-b_{2}\right\vert ^{2}$ $%
=\left( 2+\sqrt{2}\right) /4$ $\approx 0.854$,\ which indicates that at the
time $\tau =T_{rev}/4$\ the initial GWP splits into two sub-GWPs, and the
one at $N_{p0}$\ is the large partial revival.

Another example to illustrate this mechanism is when the initial state is a
superposition of two GWPs with $N_{0}=N_{A}$, $N_{B}$\ respectively, i.e.%
\begin{eqnarray}
\left\vert \Phi \left( N_{A},N_{B}\right) \right\rangle &=&\frac{1}{\sqrt{2}}%
[\left\vert \psi \left( N_{A}\right) \right\rangle +\left\vert \psi \left(
N_{B}\right) \right\rangle ]  \notag \\
&=&\sum_{k}\frac{e^{-k^{2}/2\alpha ^{2}}}{\sqrt{2\Omega _{2}}}[\sin
(kN_{A})+\sin (kN_{B})]\left\vert k\right\rangle .  \label{superp}
\end{eqnarray}%
The levels with vanishing expansion coefficients are determined by%
\begin{eqnarray}
n(N_{A}+N_{B}) &=&2Nk  \notag \\
\text{or }n(N_{B}-N_{A}) &=&(2k-1)N.
\end{eqnarray}%
One of the solutions is $N_{B}=$ $2N_{A}=$ $2N/3$. The vanishing levels are $%
n=2k$ and $n=3(2k-1)$ with $k=1,2,...$. Then the greatest common divisor is $%
24\Delta E$. The wave packet will be revived at multiples of $T_{rev}/24$
for this case, as shown in Fig. \ref{fig4}b.

\section{Numerical simulations}

The analysis above is based on the assumption that the spectrum of the
system is quadratic. However, this is only approximately true in the lower
energy range. In order to demonstrate the fractional revival in the
tight-binding system and to show how exact the approximation is, in this
section we will exhibit numerical simulations for finite size systems.

\begin{figure}[tbp]
\includegraphics[ bb=12 302 526 752, width=7 cm, clip]{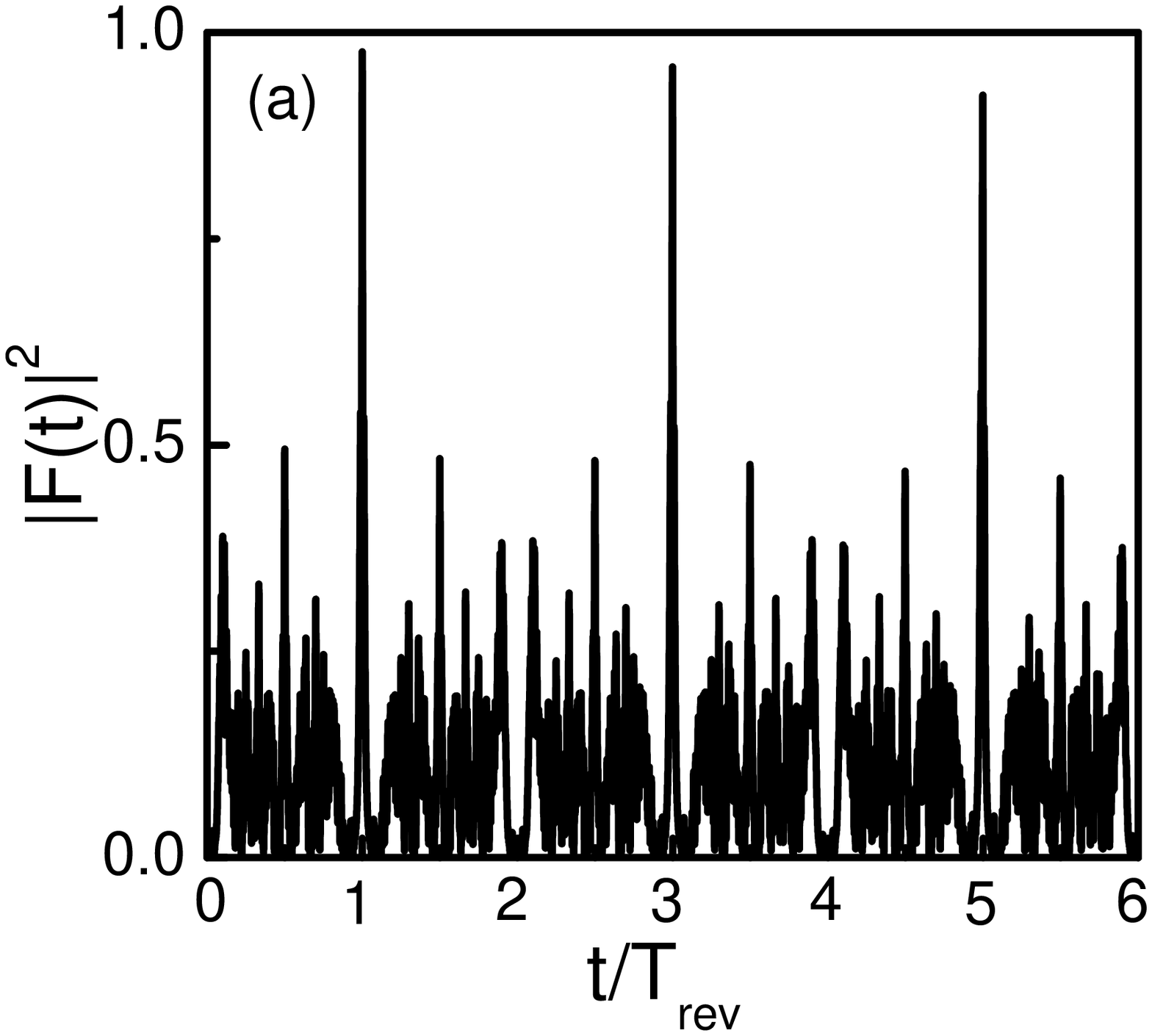} %
\includegraphics[ bb=12 302 526 752, width=7 cm, clip]{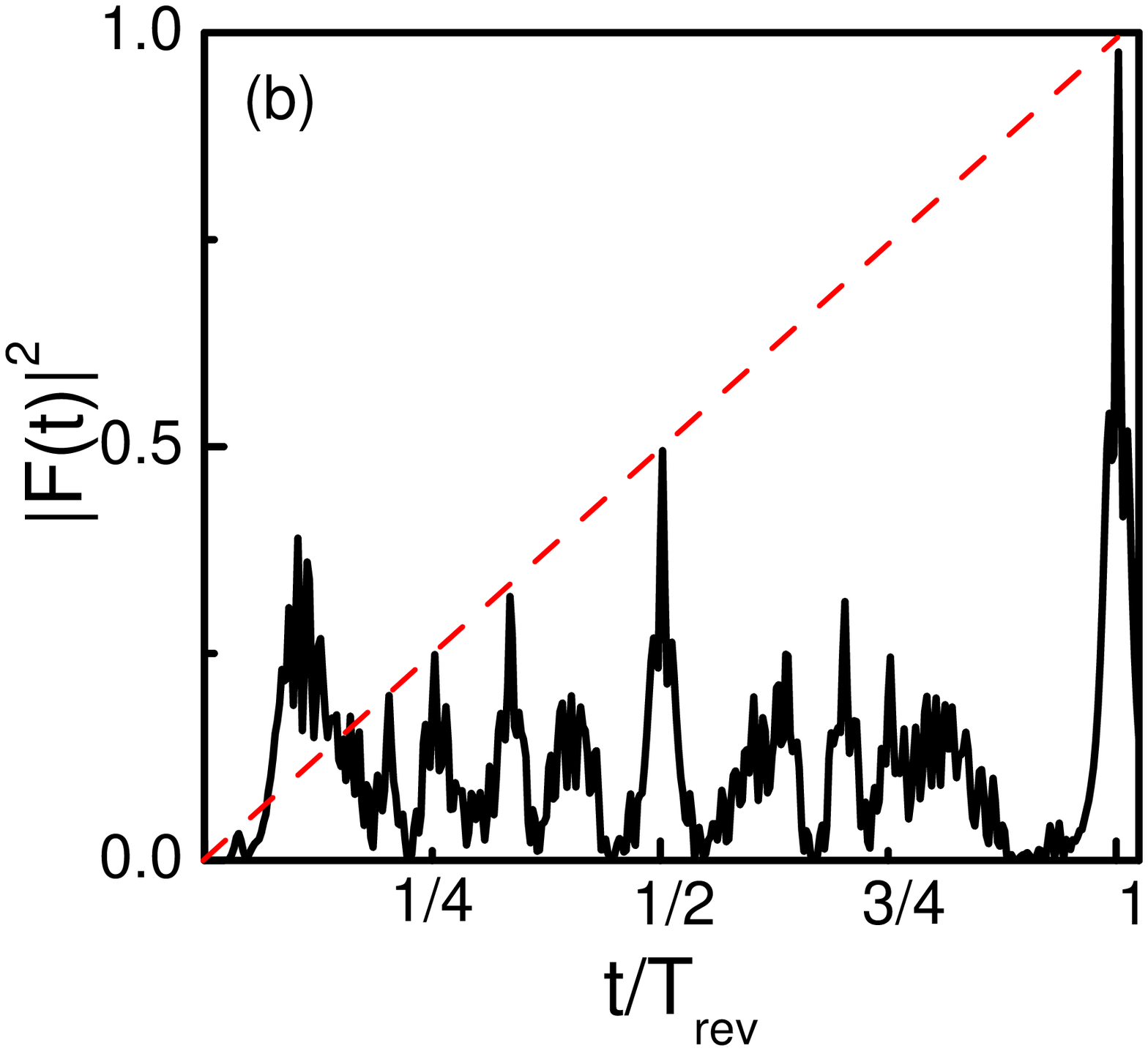}
\caption{(Color online) Plots of the square of fidelity, $\left\vert
F(t)\right\vert ^{2}$, for the initial GWP with $\Delta =24,N_{0}=50$ in the
system with $N=500$. 2(b) is a part of 2(a) over one revival time. The
dashed line indicates that the square of the fidelities at $\protect\tau %
/T_{rev}=1/2,1/3,1/4,...$ are approximately in a line.}
\label{fig2}
\end{figure}

\begin{figure}[tbp]
\includegraphics[ bb=21 96 530 755, width=8 cm, clip]{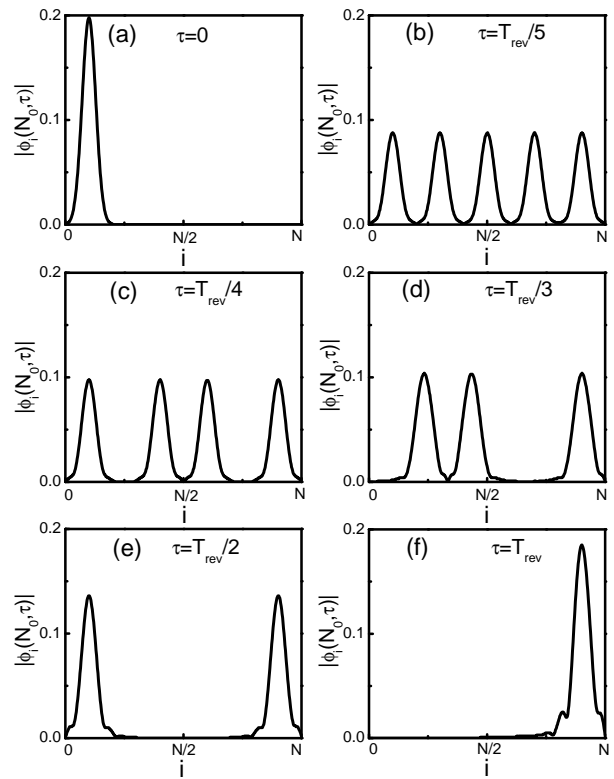}
\caption{The illustrations of the fractional revival via the time evolution
of a GWP with $\Delta =24,N_{0}=50$ in the system with $N=500$. (a) $\protect%
\tau =0$, (b) $\protect\tau =T_{rev}/5$, (c) $\protect\tau =T_{rev}/4$, (d) $%
\protect\tau =T_{rev}/3$, (e) $\protect\tau =T_{rev}/2$, (f) $\protect\tau %
=T_{rev}$. It shows that at $\protect\tau =T_{rev}p/q$, the GWP splits into
several sub-GWPs at corresponding positions with the same shape as the
initial one.}
\label{fig3}
\end{figure}

\begin{figure}[tbp]
\includegraphics[ bb=14 106 527 561, width=7.0 cm, clip]{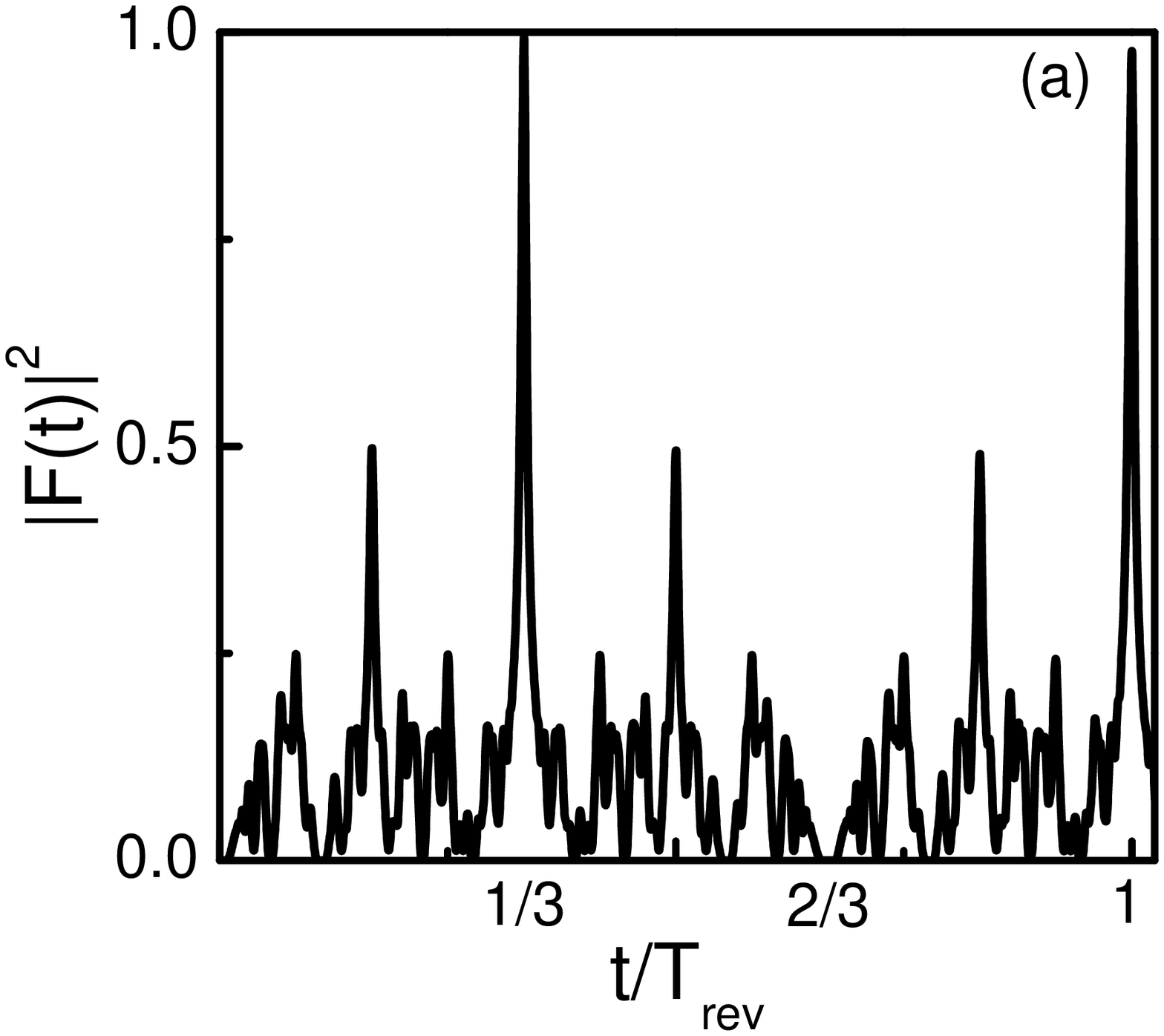} %
\includegraphics[ bb=12 62 530 528, width=7.0 cm, clip]{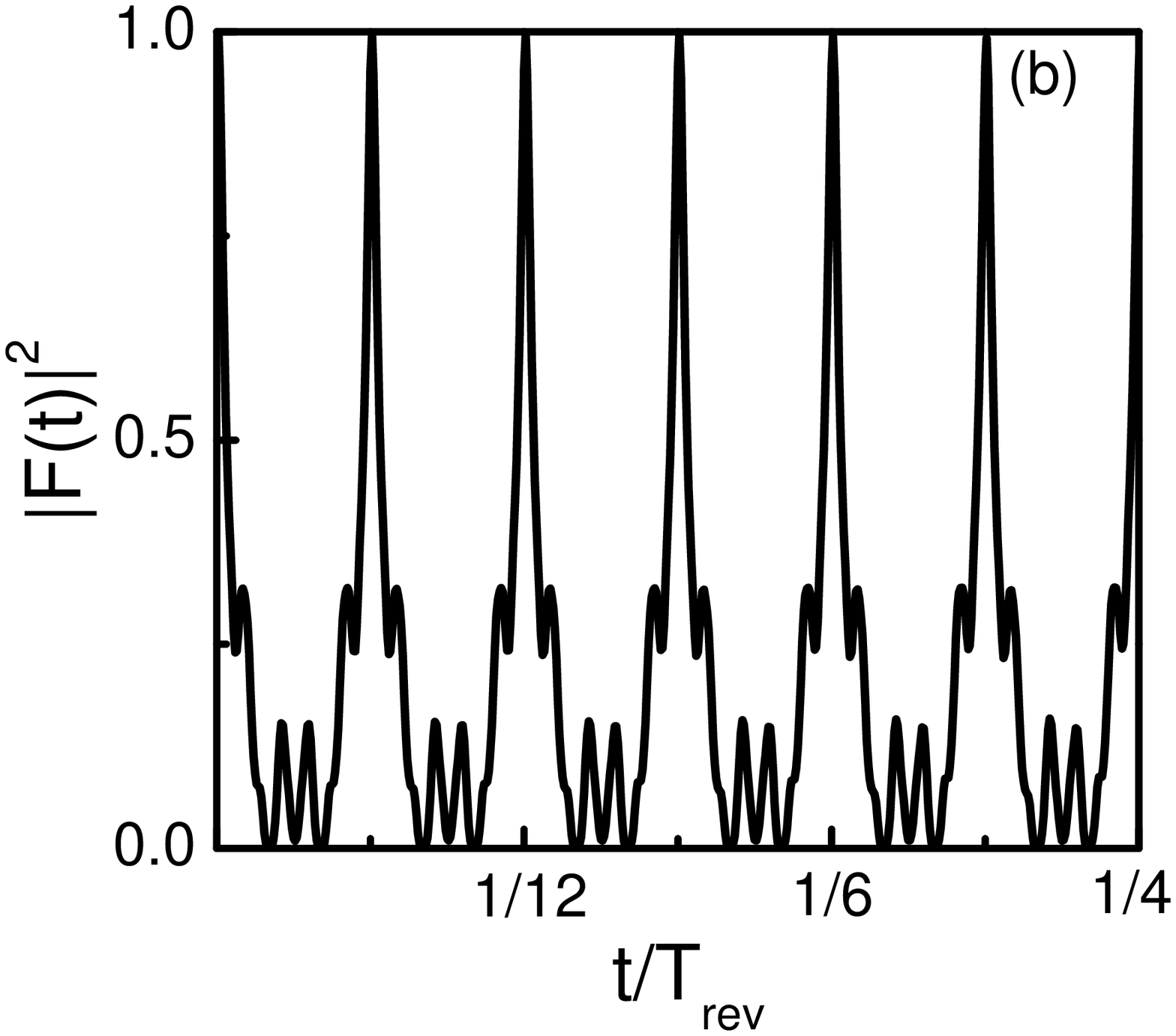}
\caption{(a) Plot of the square of fidelity $\left\vert F(t)\right\vert ^{2}$%
, same as Fig. 2b, but for $N_{0}=N/3$ over one revival time. (b) Plot of
the square of fidelity over the time $[0,T_{rev}/4]$ for the initial state
which is a superposition of two GWPs with $N_{0}=N/3,2N/3$\ respectively. It
shows that for such an initial state, the revival period is reduced.}
\label{fig4}
\end{figure}

\begin{figure}[tbp]
\includegraphics[ bb=11 304 527 752, width=7.0 cm, clip]{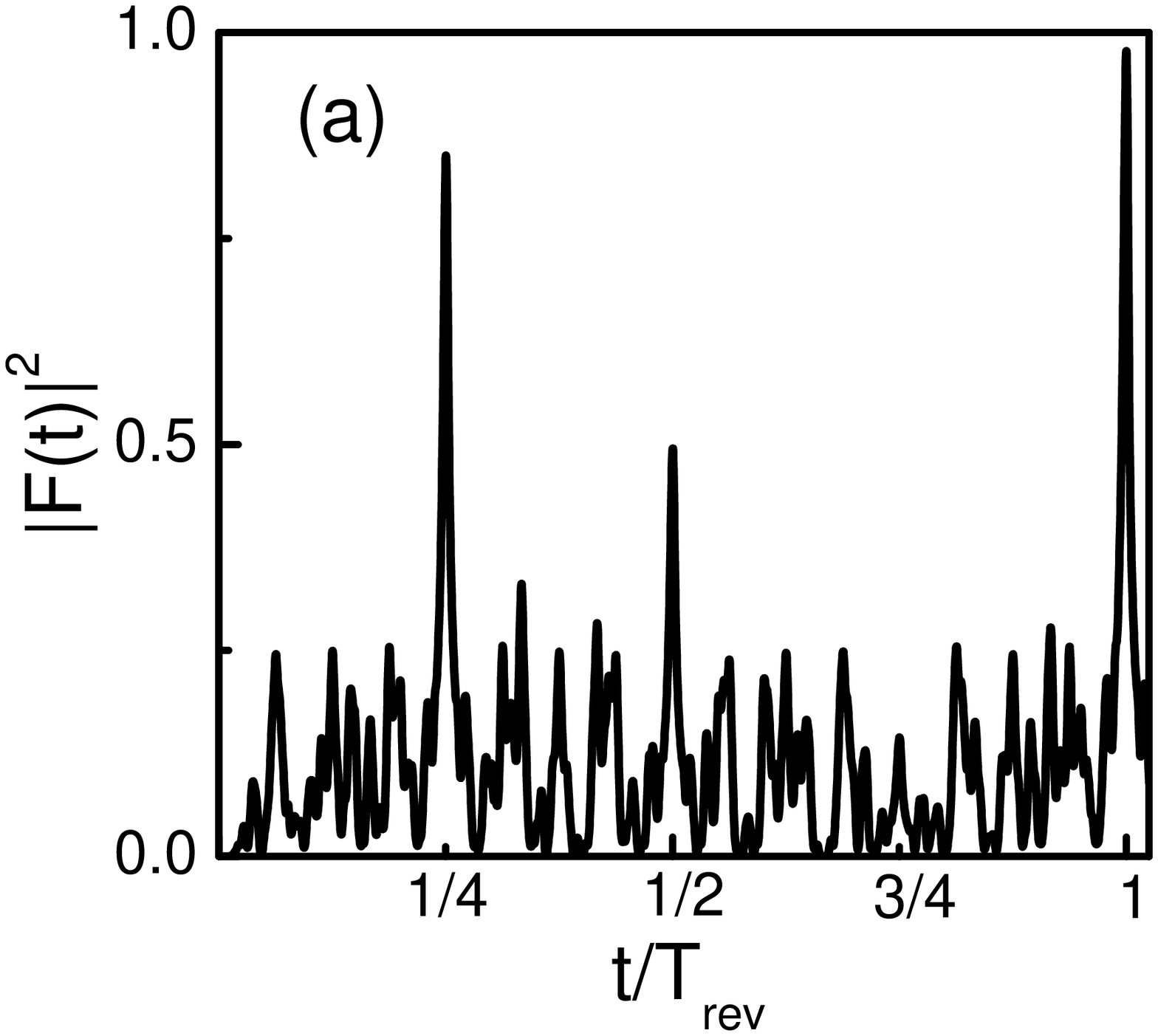} %
\includegraphics[ bb=11 307 527 752, width=7.0 cm, clip]{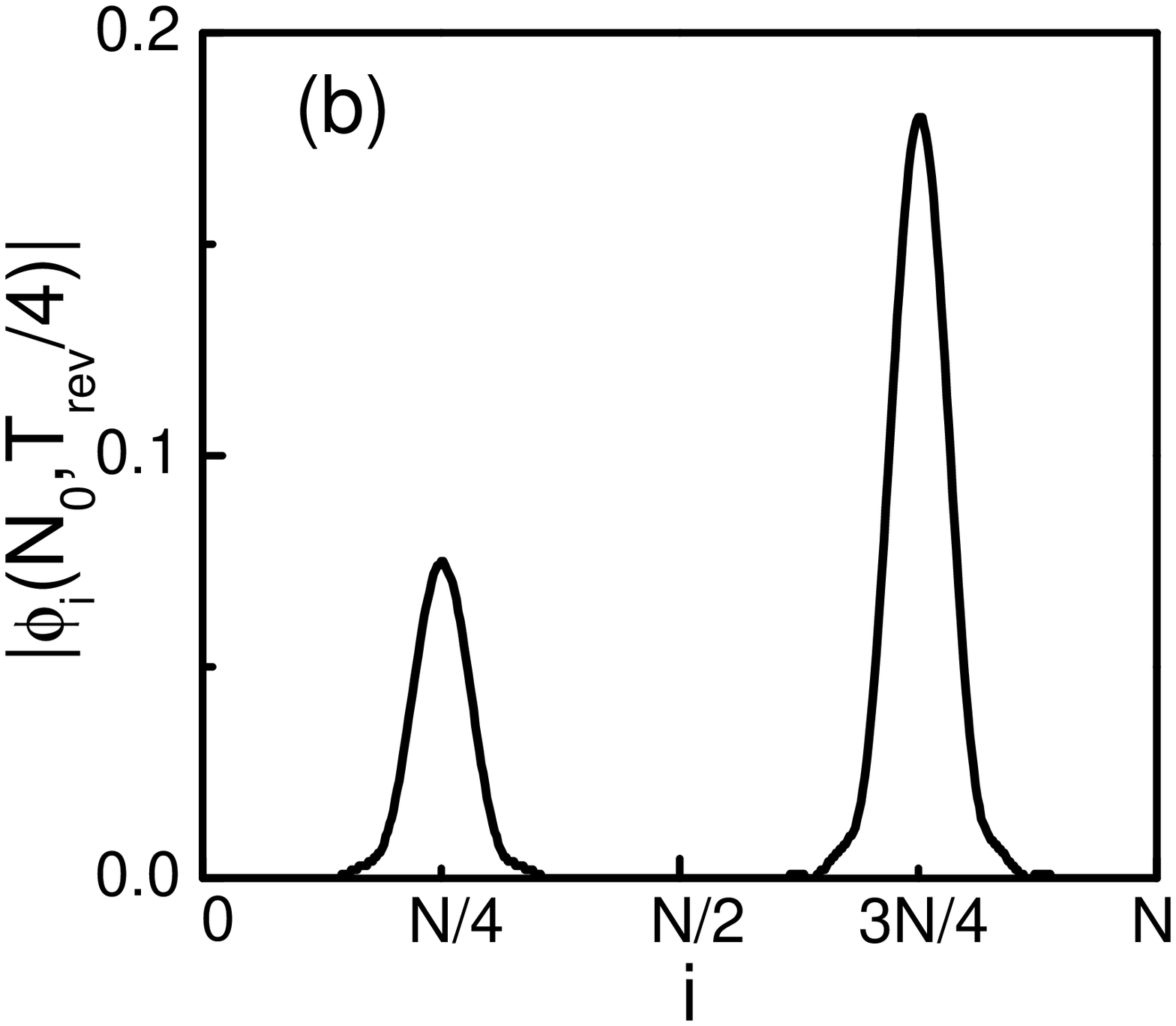}
\caption{(a) Plot of the square of fidelity $\left\vert F(t)\right\vert ^{2}$%
, same as Fig. 2b, but for $N_{0}=N/4$ over one revival time. (b) The
profile of the evolved GWP at the time $\protect\tau =T_{rev}/4$.}
\label{fig5}
\end{figure}

We start our investigation from the revival and fractional revival in the
general case. We consider the time evolution of an initial GWP with $\Delta
=24$ and $N_{0}=50$ in the system with $N=500$. According to the formalism
we have $T_{rev}=$ $\left( N+1\right) ^{2}/(2\pi J)$ $\approx 4.0\times
10^{4}/J$, which is taken as the unit of time $t$ in the numerical results.
The fidelity $\left\vert F(t)\right\vert ^{2}$\ over the interval $t\in %
\left[ 0,6T_{rev}\right] $ is plotted in Fig. \ref{fig2}a. It shows that the
fidelity has peaks around the instants $\tau =T_{rev}$, $3T_{rev}$, $%
5T_{rev} $, $...$, which is in agreement with the formalism in Sec. II.
Interestingly, there exist many regular small peaks between two neighbor big
peaks. We present the small peaks in Fig. \ref{fig2}b to show the details of
the small peaks. According to the general formalism, for local GWP, at
instants $\tau /T_{rev}=1$, $1/2$, $1/3$, $1/4$, $1/5$, $...$, the
corresponding $\left\vert F(\tau )\right\vert ^{2}$ should be equal to the
values of $\tau /T_{rev}$. The plot in Fig. \ref{fig2}b is in agreement with
the analytical results with high accuracy. In order to demonstrate the
mechanism of the fractional revival more explicitly, we calculate the
profile of the evolved state
\begin{equation}
|\phi _{i}(N_{0},\tau )|=|\left\langle i\right\vert \left. \phi (N_{0},\tau
)\right\rangle |
\end{equation}%
at $\tau /T_{rev}$ as a function of the position $i$\ and plot it in Fig. %
\ref{fig3}. From the analytical results, at instants $\tau /T_{rev}=1/5$, $%
1/4$, $1/3$, $1/2$, and $1$, the cloned sub-GWPs have the probabilities of $%
1/5$, $1/4$, $1/3$, $1/2$, and $1$, which result in the maximum values of
the corresponding sub-GWPs to be $0.089$, $0.099$, $0.114$, $0.140$ and $%
0.198$. The numerical results, the number and the shapes of the sub-GWPs are
in good accord with the theoretical prediction approximately.
%%%%%%%%%%%%%%%%%%%%%%%%%%%%%%%%%%%%%%%%%%%%%%%%%%%%%%%%%%%%%%%%%
%%%%%%%%%%%%%%%%%%%%%%%%%%%%%%%%%%%%%%%%%%%%%%%%%%%%%%%%%%%%%%%%%%
%%%%%%%%%%%%%%%%%%%%%%%%%%%%%%%%%%%%%%%%%%%%%%%%%%%%%%%%%%%%%%%%
%%%%%%%%%%%%%%%%%%%%%%%%%%%%%%%%%%%%%%%%%%%%%%%%%%%%%%%%%%%%%%%%%
Now we turn our numerical investigation to the special cases. Let us look at
a wave packet initially localized at $N_{0}=N/3$. The result introduced in
Eq. (\ref{N/3}) shows that the initial GWP recurs totally at its mirror part
at time $\tau =T_{rev}/3$. The numerical result in Fig. \ref{fig4}a shows
that the theoretical analysis is in agreement with the result of numerical
simulation represented by $\left\vert F(T_{rev}/3)\right\vert ^{2}=1$, which
has also been well explained from the SPMC. On the other hand, we also
demonstrate the evolution of $\left\vert \Phi \left( N/3,2N/3\right)
\right\rangle $ in Eq. (\ref{superp}) numerically. In Fig. \ref{fig4}b, it
shows that the first revival time is around $T_{rev}/24$ which is in
agreement with the analytical result. It also indicates that the proper
choice of initial wave packet can revive in a shorter time, which can be
used to transfer long-range entangled GWPs in the discrete system.
%%%%%%%%%%%%%%%%%%%%%%%%%%%%%%%%%%%%%%%%%%%%%%%%%%%%%%%%%%%%%%%%%%%%%%%%
%%%%%%%%%%%%%%%%%%%%%%%%%%%%%%%%%%%%%%%%%%%%%%%%%%%%%%%%%%%%%%%%%%%%%%%%%
Numerical simulation is also performed in the same system but with $%
N_{0}=N/4 $. The theoretical calculation shows that such an initial GWP
should revive at $N_{p0}=3N/4$ after the time $T_{rev}/4$\ with a relatively
higher fidelity $0.854$. The numerical results presented in Fig. \ref{fig5}a
and b depict the characteristics of the time evolution of the initial wave
packet $\left\vert \psi \left( N/4\right) \right\rangle $ via the fidelity $%
\left\vert F(t)\right\vert ^{2}$ and the profile of the wave function $|\phi
_{i}(N/4,T_{rev}/4)|$. From the analytical result of Eq. (\ref{sub}), the
behavior in Fig. \ref{fig5}a and b can be explained. Actually, at the
instant $T_{rev}/4$, the initial GWP splits into four sub-GWPs, with two of
them being at $N_{p0}$\ and two at $N_{0}$. The final shape of the wave
function should be two cloned initial wave packets with probabilities of $%
0.854$ and $0.146$ respectively. These will result in the two maxima of the
wave functions, $0.183$ and $0.076$ around the positions $3N/4$\ and $N/4$.

Based on the numerical results presented in this section, we conclude that
the fractional revival phenomena for local wave packets can be observed in
the discrete system.
%%%%%%%%%%%%%%%%%%%%%%%%%%%%%%%%%%%%%%%%%%%%%%%%%%%%%%%%%%%%%%%%%%%
%%%%%%%%%%%%%%%%%%%%%%%%%%%%%%%%%%%%%%%%%%%%%%%%%%%%%%%%%%%%%%%%%

\section{Fractional fidelity of QIT}

In general, there are two ways to employ the tight-binding model as a data
bus for quantum information transfer: (1) The qubit array is usually
described by a Heisenberg spin chain system \cite{Bose1}. Within the context
of quantum state transfer, only the dynamics of the single magnon is
relevant. Thus in the single magnon invariant subspace, this model can be
mapped into a single spinless fermion tight-binding model. The quantum
information is encoded in the superposition of the single and zero particle
states. (2) On the other hand, the quantum information can also be encoded
in the polarization of the Bloch electron (Fig. \ref{fig1}). If the spin
state of the\ Bloch electron is a conserved quantity for the Hamiltonian of
the medium, the spin state cannot be influenced during the propagation, no
matter how the spatial shape of the wave function changes \cite{YS}. In this
case, the locality of the final state is crucial. We will discuss the two
schemes in detail in the following based on the formalism of the fractional
revival in the tight-binding model.

\subsection{Scheme A: qubit array}

It is well known that, by employing the Jordan-Wigner transformation \cite%
{J-D}, the one-dimensional tight-binding chain with NN hopping is equivalent
to a simple $XY$\ chain. Such a system is usually used to depict the physics
of the qubit array. In this paper, the basis of the $XY$ chain is in the
form $\prod\nolimits_{j}\left\vert \widetilde{n}\right\rangle _{j}$\ with $%
\widetilde{n}=\widetilde{0},\widetilde{1}$, i.e., $\left\vert \widetilde{1}%
\right\rangle _{j}=\left\vert \uparrow \right\rangle _{j}$\ and $\left\vert
\widetilde{0}\right\rangle _{j}=\left\vert \downarrow \right\rangle _{j}$.
In general, the transmission of a qubit state from the location $A$ to $B$
can be regarded as the following process. The initial qubit state $%
\left\vert \psi _{A}\right\rangle =u\left\vert \widetilde{1}\right\rangle
_{A}+v\left\vert \widetilde{0}\right\rangle _{A}$ is prepared at $A$. If one
can find an operation $U_{AB}$\ to realize
\begin{equation}
U_{AB}\left\vert \psi _{A}\right\rangle =u\left\vert \widetilde{1}%
\right\rangle _{B}+e^{i\varphi _{AB}}v\left\vert \widetilde{0}\right\rangle
_{B},
\end{equation}%
where $\varphi _{AB}$\ is the known phase for a given system, we say that
the qubit state is transferred from $A$ to $B$ perfectly. Then the perfect
QIT is accomplished. Bose \cite{Bose1} proposed that the operation $U_{AB}$\
can be achieved in the qubit array by the time evolution of the system based
on the fact that the saturated ferromagnetic state $\prod\nolimits_{j}\left%
\vert \widetilde{0}\right\rangle _{j}$ is an eigenstate of the model.
Furthermore, it is found that perfect state transfer can be implemented if
the system meets the SPMC. Unfortunately, for an array to satisfy the SPMC,
it requires modulation \cite{Christandle1,Christandle2,Christandle3, Shitao2}
of the couplings between qubits, which is difficult to pre-engineer in
experiments.

Nevertheless, we can consider the transferred state to be not a single-qubit
state at a certain site but a superposition of single-qubit states localized
in a small range of the coordinate space. Together with the saturated
ferromagnetic state, the quantum information can be encoded in such a single
magnon Gaussian wave packet at $N_{0}$
\begin{equation}
\left\vert \psi _{N_{0}}\right\rangle =\frac{1}{\sqrt{\Omega _{1}}}%
\sum_{i}e^{-\alpha ^{2}(i-N_{0})^{2}/2}(u\left\vert \widetilde{1}%
\right\rangle _{i}+v\left\vert \widetilde{0}\right\rangle
_{i})\prod\limits_{j\neq i}\left\vert \widetilde{0}\right\rangle _{j}.
\label{gwp}
\end{equation}%
This state contains the same quantum information as that of the single qubit
state $\left\vert \psi _{A}\right\rangle $. So if the GWP (\ref{gwp})
appears completely at another place, perfect QIT is accomplished. In the
following, we will show that if the GWP (\ref{gwp}) appears partially,
perfect QIT can also be accomplished.

According to the formalism of the fractional revival, we note that at $\tau
=pT/q$ there always exists a cloned sub-GWP of the initial state at the
mirror position, with the probability $\left\vert b_{l/2}\right\vert ^{2}$.
From the point of view of quantum information, theoretically, the
information of initial state encoded in the initial state by factors $u$ and
$v$ has been transferred to its counterpart completely, although the
fidelity of QST $F(\tau )$ may be far from $1$.

In order to depict this fact, we introduce the \textit{fractional fidelity, }%
which is expressed as%
\begin{eqnarray}
\left\vert F_{f}(\tau )\right\vert &=&\frac{1}{\left\vert b_{l/2}\right\vert
}\left\vert F(\tau )\right\vert  \label{ff} \\
&=&\frac{1}{\left\vert b_{l/2}\right\vert }\left\vert
-b_{l/2}+\sum_{r=1,\lambda =\pm }^{l/2-1}b_{r}f(N_{p0},N_{r}^{\lambda
})\right\vert ,  \notag
\end{eqnarray}%
and is unity if the retrieved sub-GWP is the exact clone of the initial
state. Then the QIT can be transferred perfectly, even the QST is not
completely.%
%%%%%%%%%%%%%%%%%%%%%%%%%%%%%%%%%%%%%%%%%%%%%%%%%%%%%%%%%%%%%%%%%
\begin{figure}[tbp]
\includegraphics[ bb=15 305 524 753, width=7 cm, clip]{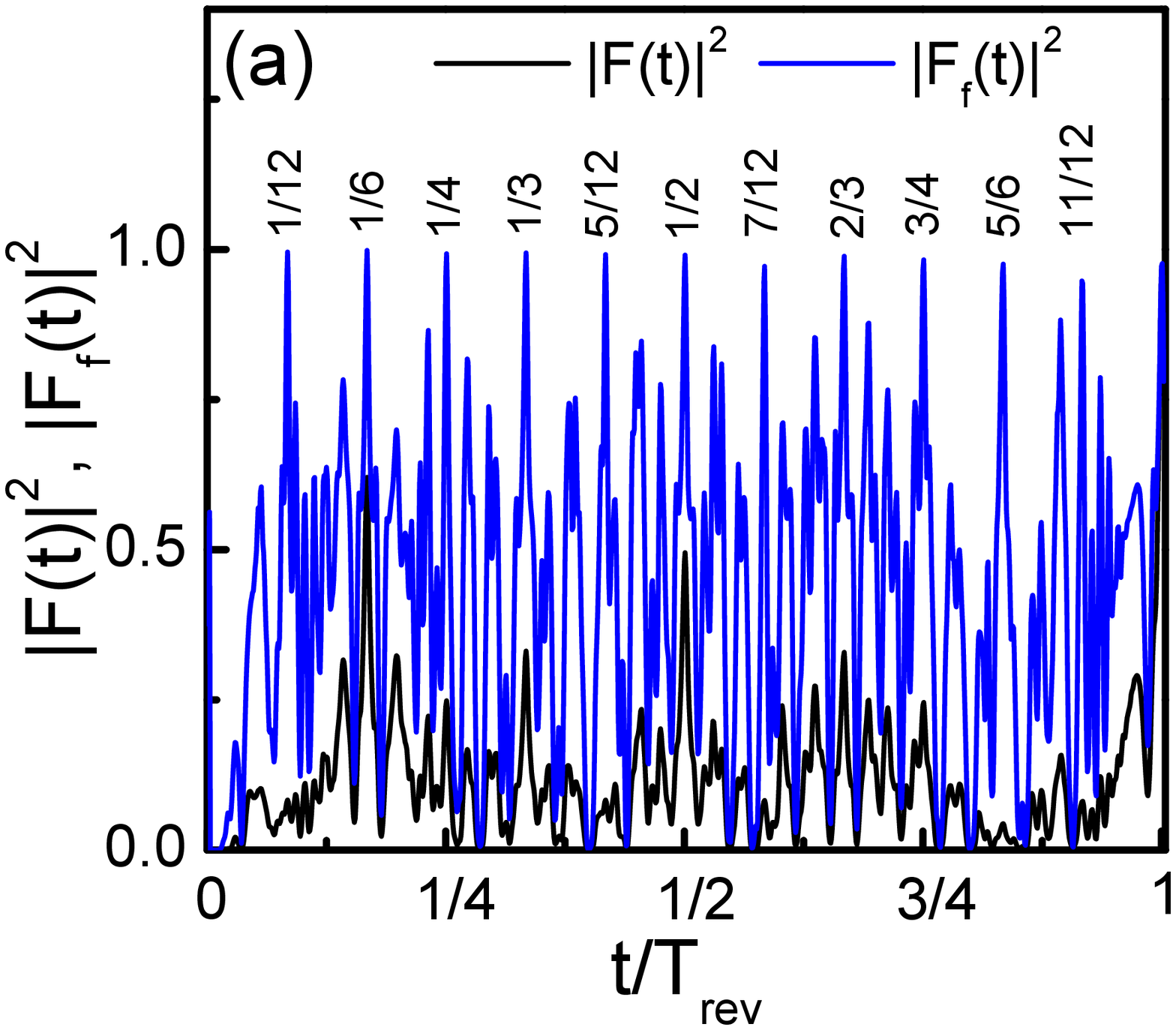} %
\includegraphics[ bb=15 305 524 753, width=7 cm, clip]{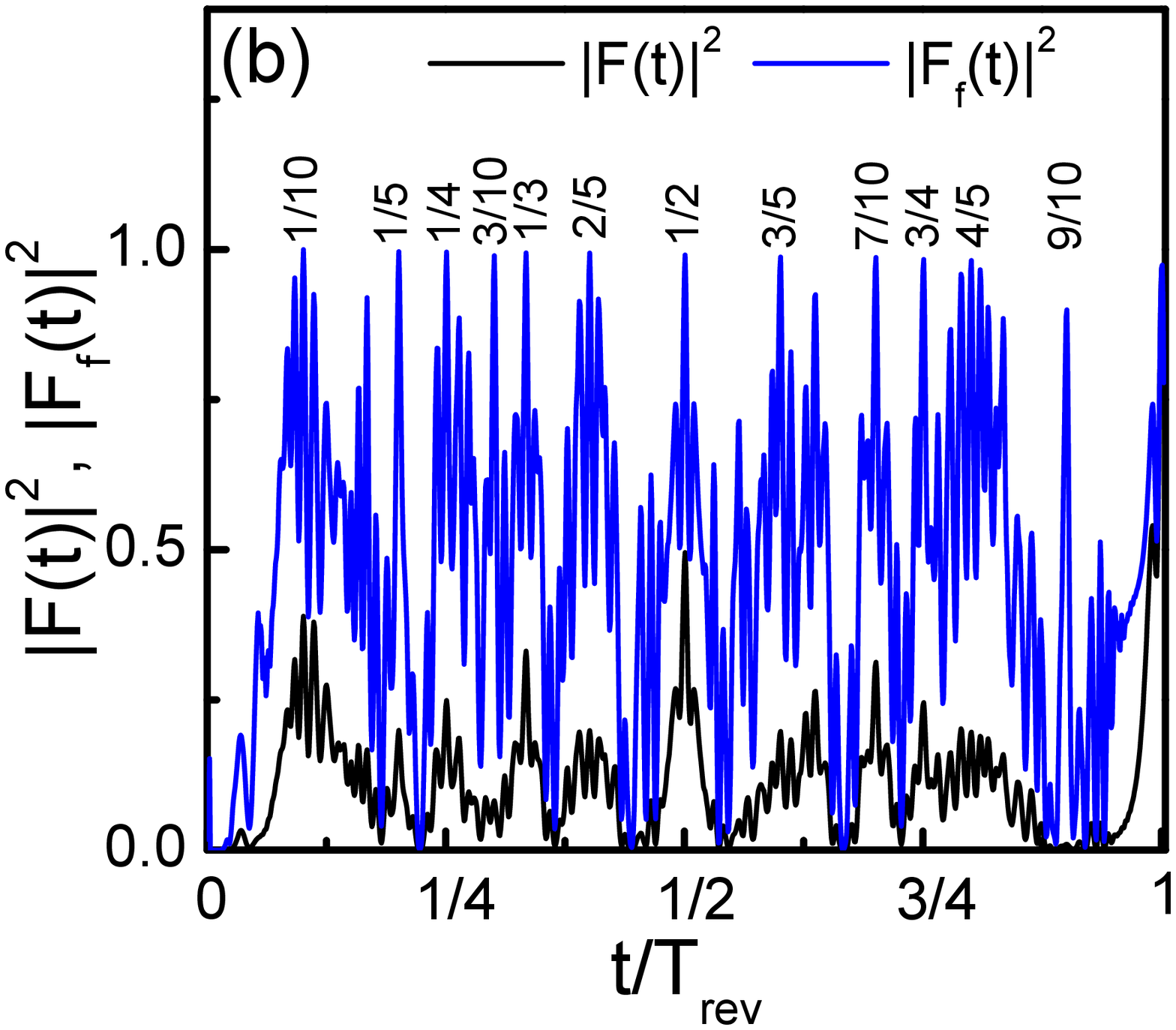}
\caption{(Color online) Plot of the square of the fractional fidelity $%
\left\vert F_{f}(t)\right\vert ^{2}$ for the GWP over one revival time with $%
\Delta =24$ in the system with $N=500$ for (a) $N_{0}=N/6$, (b) $N_{0}=N/10$%
. It shows that $\left\vert F_{f}(t)\right\vert ^{2}$ is close to $1$ at
several specified instants.}
\label{fig6}
\end{figure}
%%%%%%%%%%%%%%%%%%%%%%%%%%%%%%%%%%%%%%%%%%%%%%%%%%%%%%%%%%%%%%%%%
\begin{figure}[tbp]
\includegraphics[ bb=11 305 524 753, width=7 cm, clip]{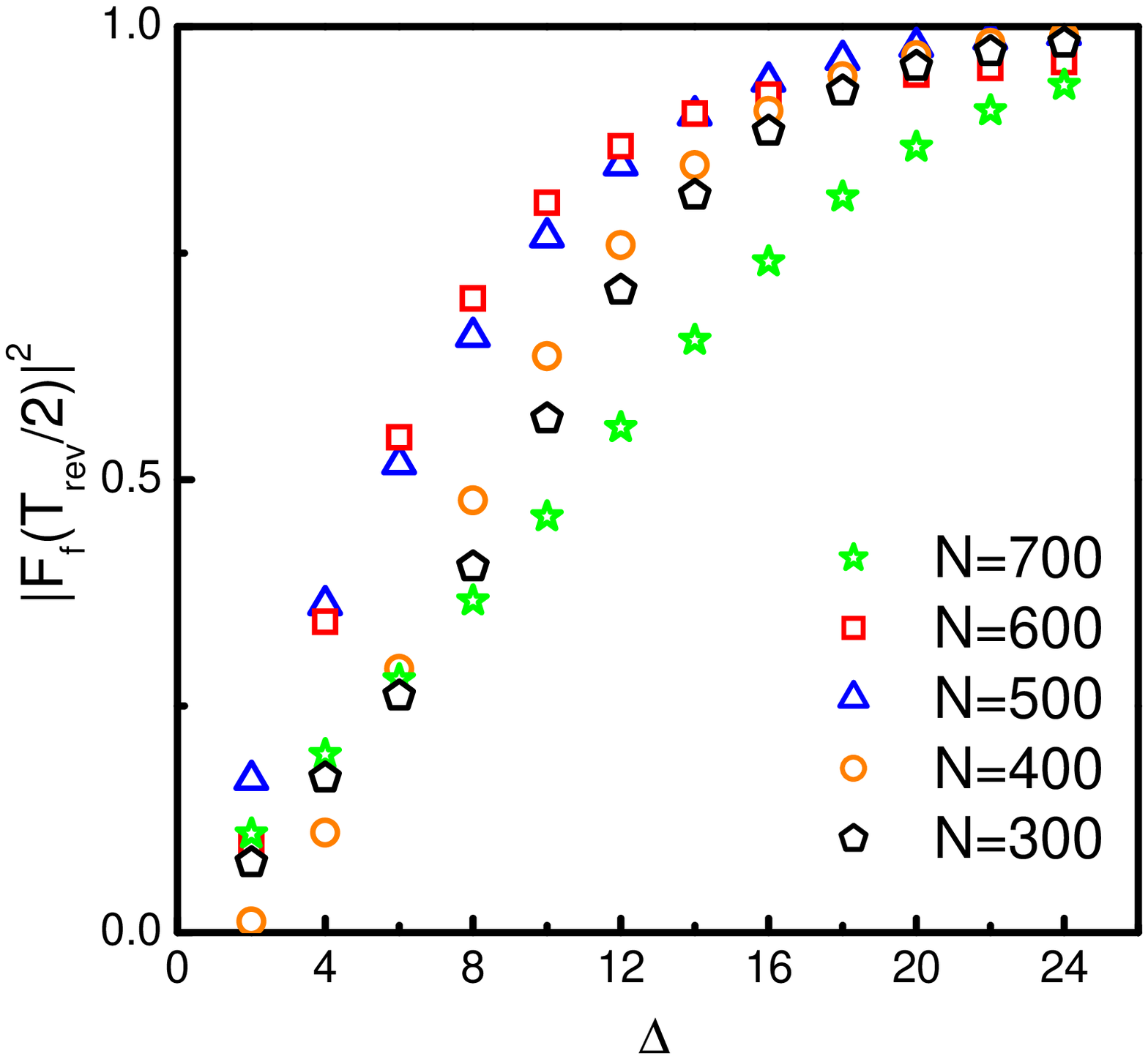}
\caption{(Color online) The square of fractional fidelity $\left\vert
F_{f}(T_{rev}/2)\right\vert ^{2} $ versus $\Delta $ for systems with various
lengths $N\in \lbrack 300,700]$. It shows that enlarging $\Delta $ can
enhance the fractional fidelity.}
\label{fig7}
\end{figure}
%%%%%%%%%%%%%%%%%%%%%%%%%%%%%%%%%%%%%%%%%%%%%%%%%%%%%%%%%%%%%%%%%

To demonstrate the new definition of the fidelity and verify how much the
initial state and the sub-GWP are alike, numerical simulation is performed
in a finite system. The square of the fractional fidelity $\left\vert
F_{f}(t)\right\vert ^{2}$ for the propagation of a GWP with $\Delta =24$
from $N_{0}=N/6,$ $N/10$ to $N_{p0}=5N/6,$ $9N/10$ in the system with $N=500$
within $[0,T_{rev}]$ is plotted in Fig. \ref{fig6}. It is obvious that at
many instants $\tau =T_{rev}p/q$, the square of the fractional fidelity
approximately equal to $1$, while $\left\vert F_{f}(\tau )\right\vert ^{2}$
is still far from $1$ for many possible $T_{rev}p/q$. We also notice that
the initial position affects the results strongly. These phenomena can be
explained as follows.

As discussed in Sec. III (A), the final state should be the superposition of
the $l$ cloned sub-GWPs. Although there always exists a cloned sub-GWP at
the position $N_{p0}$, the overlap of these sub-GWPs affects $\left\vert
F_{f}(\tau )\right\vert $. If the $l$ cloned sub-GWPs are well separated,
i.e., $f(N_{p0},N_{r}^{\pm })=0$, then we have $\left\vert F_{f}(\tau
)\right\vert =1$ directly from the Eq. (\ref{ff}). On the other hand, the
orthogonality of the $l$ cloned sub-GWPs is determined by their number and
positions, which depend on the factors $q$ and $N_{0}$. Obviously the
smaller $l$\ is, the more perfect the fractional revival is. However, in the
case that the initial position $N_{0}$\ and $l$\ satisfy the condition

\begin{equation}
(N+1-2N_{0})l=\text{integer}\times 2\left( N+1\right)  \label{commensurate}
\end{equation}%
the fractional revival should be still perfect. The Eq. (\ref{commensurate})
guarantees that the final state consists of several well separated cloned
GWPs. This is in agreement with the numerical simulations. Fig. \ref{fig6}a
shows that $\left\vert F_{f}(\tau )\right\vert ^{2}\sim 1$\ at $\tau
=T_{rev}p/12$ $(p=$ $1,2,\cdots ,11)$, while it occurs at $\tau =T_{rev}p/10$
$(p=$ $1,2,\cdots ,9)$\ in Fig. \ref{fig6}b.

\subsection{Scheme B: Flying qubit}

There is another scheme for QIT when the spin degree of freedom is
considered. We can imagine an electronic wave packet with spin polarization
as an analog of a photon \textquotedblleft flying qubit\textquotedblright ,
i.e. a polarized photon where the quantum information is encoded in its two
polarization states. We define the solid-state \textquotedblleft flying
qubit\textquotedblright , at the location $A$ in a quantum wire, as the
superposition of two orthogonal Bloch electronic GWPs $\left\vert \uparrow
\right\rangle _{A}$ and $\left\vert \downarrow \right\rangle _{A}$, where%
\begin{equation}
\left\vert \sigma =\uparrow ,\downarrow \right\rangle _{A}=\frac{1}{\sqrt{%
\Omega _{1}}}\sum_{j}e^{-\alpha ^{2}(j-N_{A})^{2}/2}c_{j,\sigma }^{\dag
}\left\vert 0\right\rangle .  \label{flying qubit}
\end{equation}%
Obviously, the two orthogonal states evolve independently. Then an arbitrary
state
\begin{equation}
\left\vert \Psi (0)\right\rangle _{A}=u\left\vert \uparrow \right\rangle
_{A}+v\left\vert \downarrow \right\rangle _{A}
\end{equation}%
will evolve to
\begin{equation}
\left\vert \Psi (t)\right\rangle =u\left\vert \phi (\uparrow
,t)\right\rangle +v\left\vert \phi (\downarrow ,t)\right\rangle ,
\end{equation}%
where $\left\vert \phi (\sigma ,t)\right\rangle =\sum_{j}f(j,t)c_{j,\sigma
}^{\dag }\left\vert 0\right\rangle $\ with $f(j,t)$\ being a
spin-independent function. The quantum information encoded in the spin state
of the initial state (\ref{flying qubit}) is carried along by the electron
and unaffected by the transfer. Therefore, the initial state will be
transferred to another location if $f(j,t)$\ is known and is still a
localized function. Of course, the simplest case is that $f(j,t)$\ is a
mirror or translation of the initial GWP \cite{YS}. In fact, if $f(j,t)$\ is
partially local at several places, the quantum information can be
transported to multiple receivers. This fact indicates that such a system
can be used for a \textquotedblleft quantum fanout\textquotedblright , which
was recently proposed by A.D. Greentree, S.J. Devitt, and L.C.L. Hollenberg
\cite{multi}. In our work, we only employ a simple open chain without any
dynamic control.\textbf{\ }In this sense, it acts as a solid-state based
splitter, entangler \cite{YS2}, and quantum fanout.

\subsection{Validity of the schemes}

In Sec. III, the analytical conclusion is only valid for lower energy GWPs.
For an arbitrary GWP, the factor $\alpha $\ determines the behavior of the
final state as time evolves. As pointed above, for the GWP with narrow width
at $k\sim 0$\ in momentum space, the effective dispersion relation is
quadratic approximately. Then the width $\Delta $\ of GWP in real space
should affect the fractional fidelity of the fractional revival. Numerical
simulation was employed to investigate the relationship between the
fractional fidelity $F_{f}(\tau =T_{rev}/2)$ and $\Delta $ with the GWP
transferring from $N_{0}=50$ to $N_{0p}=N-50$. The numerical results for $%
N=300$, $400$, $500$, $600$ and $700\ $are plotted in Fig. \ref{fig7}. It
shows that when $\Delta $\ tends to $24$, $F_{f}(\tau =T_{rev}/2)\sim 1$ for
different sizes of system. Thus it indicates that when fractional revivals
in the discrete system are employed for quantum information transmission,
the width of the chosen GWP should be more than $24$ times the lattice
spacing.

\section{Summary and discussion}

In summary, we have studied the phenomenon of fractional revivals in a
discrete system by theoretical analysis and numerical simulations of the
evolution of a GWP in a tight-binding model. It is found that, for a proper
chosen initial state, its fractional revival states have the same formalism
as that in the infinite square well. On the other hand, numerical
simulations show that the formulas of the theoretical analysis are very
accurate for the GWP.

We also proposed the concept of the fractional fidelity $F_{f}(t)$ when the
fractional revival phenomenon is exploited to achieve QST in the solid-state
system. We showed that the fractional fidelity approximately equals to $1$
at many instants. With an appropriately chosen width, the GWP with a
polarized spin state can be regarded as a flying qubit in the solid-state
system to implement quantum information transmission.

It is worthwhile to discuss the applicability of the scheme presented above.
Experimentally, the tight-binding model can be realized by a quantum dot
array, SQUID array, etc.. In a real system, quantum decoherence is the main
obstacle to the experimental implementation of quantum information. In our
scheme for quantum state transfer, the quantum decoherence time limits the
scale of the quantum wire. For coupled quantum dots, experiments show that
the coupling strength $J\sim 10$\ $meV$\ \cite{coupling constant}.
Therefore, for a $N$-site chain, the revival period $T_{rev}=(N+1)^{2}/(J\pi
)$ $\approx 1.6\times 10^{-11}N^{2}$ $ms$. On the other hand, the
decoherence time of a quantum dot is $\tau \sim 1ms$\ \cite{Decoherence time}%
. For example, considering a quantum dot array with $N=500$, we have $%
T_{rev}\approx $ $4\times 10^{-6}ms$, which is much smaller than $\tau $.
Moreover, if we perform $n$\ times full revival within the decoherence time $%
\tau $, the maximal size is $2.5\times 10^{5}/\sqrt{n}$. Therefore, for $%
n\sim 10^{4}$, the size of the system should be limited to $10^{3}$, which
implies the applicability of the scheme in practice.

We acknowledge the support of the CNSF (grant No. 90203018, 10474104,
10447133), the Knowledge Innovation Program (KIP) of Chinese Academy of
Sciences, the National Fundamental Research Program of China (No.
2001CB309310).

\end{document}